\documentclass[twocolumn,aps,prl,superscriptaddress,floatfix]{revtex4-1}
\usepackage{graphicx,subfigure}
\usepackage{amsmath,amssymb,bm}
\usepackage{mathtools}
\usepackage{multirow}
\usepackage{makecell}
\usepackage{textcomp}
\usepackage{float}
\usepackage{color}
\usepackage[normalem]{ulem}
\usepackage{dsfont}
\usepackage{mathrsfs}
\usepackage{braket}
\usepackage{transparent}
\usepackage{xcolor}
\usepackage{hhline}
\usepackage{graphicx}
\usepackage{pdfpages}
\usepackage[export]{adjustbox}
\usepackage{hyperref}
\usepackage{physics}

\makeatletter
\AtBeginDocument{\let\LS@rot\@undefined}
\makeatother

\newcommand{\figone}{
 \begin{figure*}[hbt]
    \centering
    \includegraphics[width=0.85\textwidth]{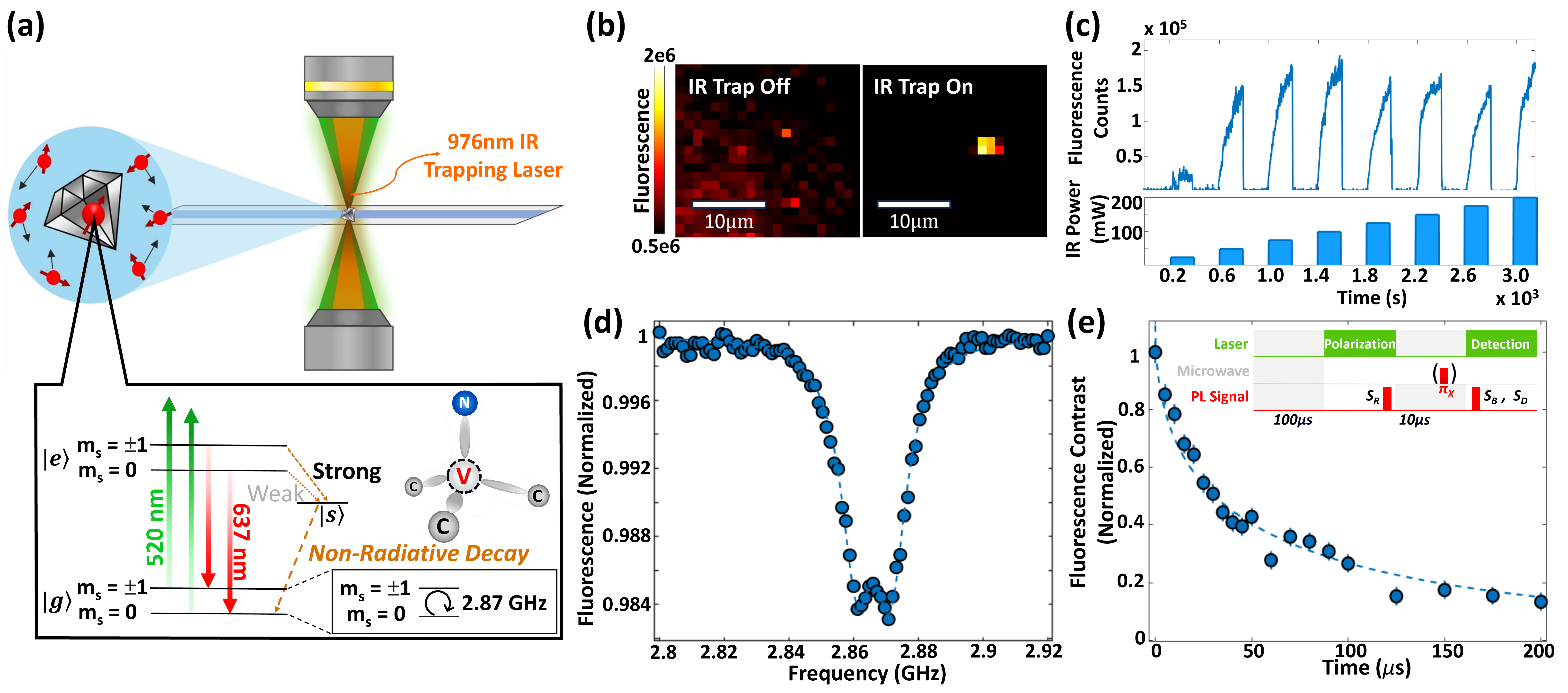}
    \caption{{\bf Optically Detected Magnetic Resonance (ODMR) spectra and \(T_1\) measurements on optically-trapped nanodiamonds}
    (a) Schematic of the experiment with NV centers in optically-trapped fluorescent nanodiamond. Bottom diagram: the lattice structure and energy levels of an NV center including a triplet ground state and excited state. 
    (b) Confocal fluorescence scan of FNDs in deionized water with and without an infrared optical trapping laser.
    (c) Fluorescence counts of the optically-trapped FNDs with increasing trapping laser power.
    (d) Measured ODMR spectrum of NV centers from optically-trapped FNDs in the absence of a magentic field. The dashed line corresponds to the fit using the sum of two Lorentzians.
    (e) Measured spin relaxation ($T_1$) dynamics of NV centers from optically-trapped $70~$nm FNDs. The dashed line is the fit using a stretched exponential decay model. Inset: Differential measurement pulse sequence for the $T_1$ experiment. Error bars represent 1 s.d. accounting statistical uncertainties.
    \label{fig:fig1}}
\end{figure*}
}

\newcommand{\figtwo}{
 \begin{figure}[hbt]
    \centering
    \includegraphics[width=0.47\textwidth]{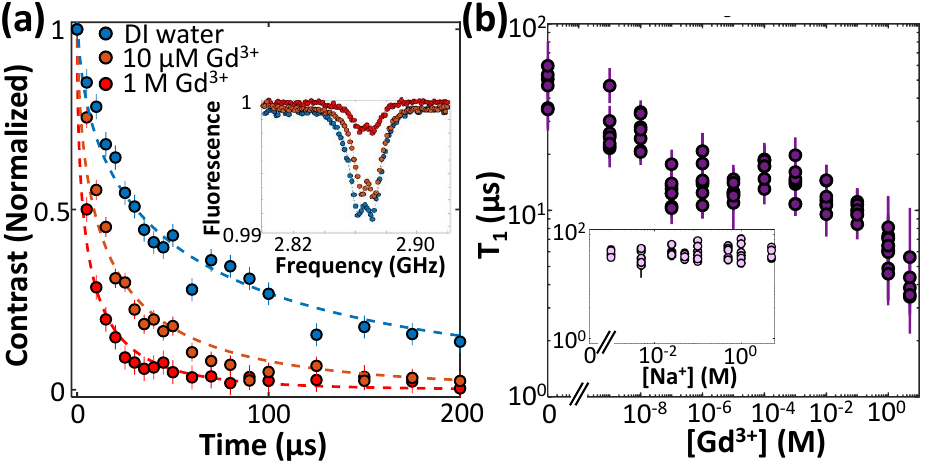}
  
    \caption{{\bf Dependence of NV relaxation time, $T_1$, with the concentration of paramagnetic species in deionized water}
    (a)  Measured NV spin relaxation dynamics from optically-trapped FNDs in deionized water, $10~\mu$M \(\mathrm{Gd}^{3+}\), and 1~M \(\mathrm{Gd}^{3+}\) solutions. Dashed lines correspond to the fits with stretched exponential decays. Inset: Measured NV ODMR spectra in these solutions. Dashed lines correspond to the fits using the sum of two Lorentzians. Error bars represent 1 s.d. accounting statistical uncertainties.
    (b) The measured NV $T_1$ dependence on \(\mathrm{Gd}^{3+}\) concentration. Inset: The measured NV $T_1$ dependence on \(\mathrm{Na}^{+}\) concentration as a reference. Error bars in time represent 1 s.d. accounting fitting error.
    }
    \label{fig:fig2}
\end{figure}
}

\newcommand{\figthree}{
 \begin{figure*}[t]
    \centering
    \includegraphics[width=0.9\textwidth]{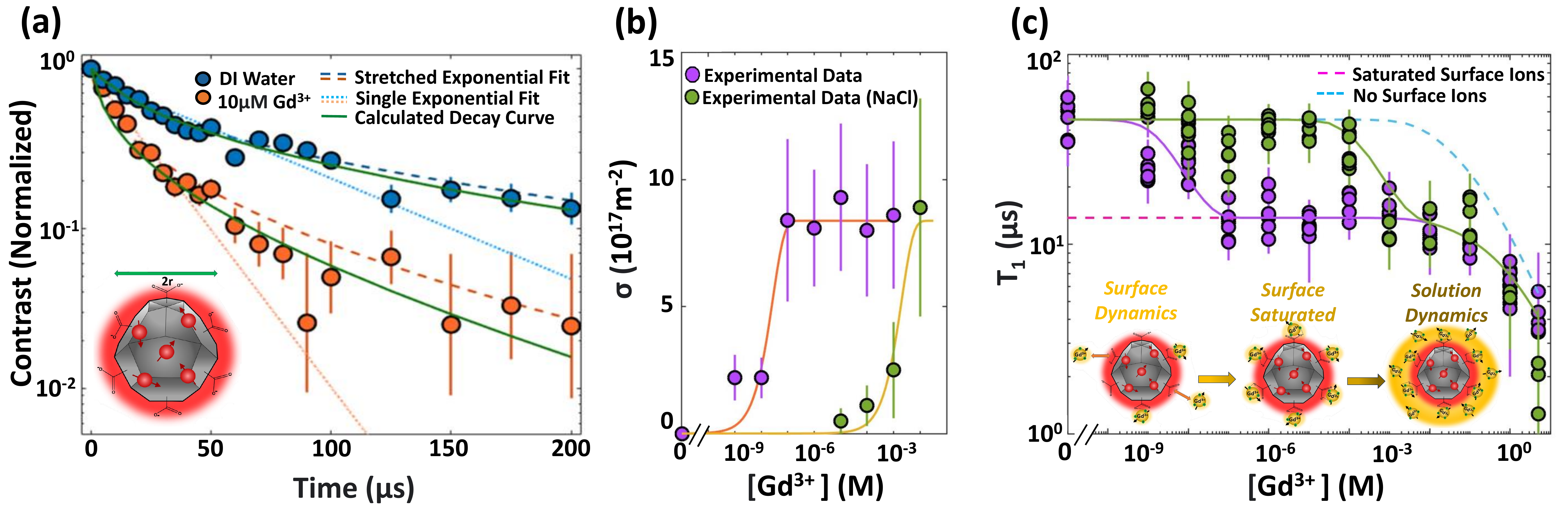}
    \caption{{\bf Theoretical model for NV relaxation dynamics in FNDs in the presence of paramagnetic spins}
    (a) Measured NV spin relaxation dynamics from optically-trapped FNDs in deionized water and $10~\mu$M \(\mathrm{Gd}^{3+}\). 
    The solid green lines are the numerically calculated decay curves using our model. The darker dashed lines are the fits using stretched exponential decay, $\sim e^{-(t/T_1)^{0.5}}$, while the lighter dotted lines correspond to the fits with single exponential profiles. The contrast is plotted in logarithmic scale for a better comparison between the two decay profiles. Inset: Schematic diagram showing NV centers at different locations inside a FND with radius $r$. The surface of FND is coated with a layer of negatively charged chemical structures that can attract positive ions from the solution. Error bars represent 1 s.d. accounting statistical uncertainties.
    (b) The fitted surface $\mathrm{Gd}^{3+}$ concentration, $\sigma$, at different solution concentration, $\rho$. The solid lines correspond to the fit from our model \cite{supp}. Error bars in time represent 1 s.d. accounting statistical uncertainties.
  (c) The measured NV $T_1$ dependence with \(\mathrm{Gd}^{3+}\) concentration. The solid lines correspond to the calculated curve from our model, containing both spins on the FND surface and in the solution. The dashed pink line corresponds to the assumption of a fixed surface concentration; the dashed blue line corresponds to not considering the effects of ions on NV $T_1$. Our model faithfully captures all three different phases of the measured $T_1$ data, i.e. surface dynamics, surface saturation and solution dynamics. Insets: schematics showing the FND configuration at the three different regimes. Error bars in time represent 1 s.d. accounting fitting error.
    }
    \label{fig:fig3}
\end{figure*}
}

\usepackage{graphicx}

\graphicspath{ {figures/} }

\makeatletter
\let\saved@includegraphics\includegraphics
\AtBeginDocument{\let\includegraphics\saved@includegraphics}
\makeatother

\makeatletter
\newcommand*{\centerfloat}{%
  \parindent \z@
  \leftskip \z@ \@plus 1fil \@minus \textwidth
  \rightskip\leftskip
  \parfillskip \z@skip}
\makeatother

\usepackage{lineno} 
\begin{document}

\title{Optically-Trapped Nanodiamond-Relaxometry Detection of Nanomolar Paramagnetic Spins in Aqueous Environments}

\author{
Shiva~Iyer,$^{1,2,*}$
Changyu~Yao,$^{1,*}$
Olivia Lazorik,$^{1,*}$
Md Shakil Bin Kashem,$^{1}$
Pengyun Wang,$^{1}$
Gianna Glenn,$^{1}$\\
Michael Mohs,$^{1}$
Yinyao Shi,$^{1}$
Michael Mansour,$^{1}$
Erik Henriksen,$^{1,2,3}$
Kater Murch,$^{1,2}$\\
Shankar Mukherji,$^{1,2,4,5,^\dag}$
Chong Zu$^{1,2,3,^\dag}$
\\
\medskip
\normalsize{$^{1}$Department of Physics, Washington University, St. Louis, MO 63130, USA}\\
\normalsize{$^{2}$Center for Quantum Leaps, Washington University, St. Louis, MO 63130, USA}\\
\normalsize{$^{3}$Institute of Materials Science and Engineering, Washington University, St. Louis, MO 63130, USA}\\
\normalsize{$^{4}$Department of Cell Biology and Physiology, Washington University School of Medicine, St. Louis, MO 63130, USA}\\
\normalsize{$^{5}$Center for Biomolecular Condensates, Washington University, St. Louis, MO 63130, USA}\\
\normalsize{$^*$These authors contributed equally to this work}\\
\normalsize{$^\dag$To whom correspondence should be addressed; E-mail: smukherji@wustl.edu, zu@wustl.edu}\\
}

\begin{abstract}
Probing electrical and magnetic properties in aqueous environments remains a frontier challenge in nanoscale sensing. Our inability to do so with quantitative accuracy imposes severe limitations, for example, on our understanding of the ionic environments in a diverse array of systems, ranging from novel materials to the living cell.
The Nitrogen-Vacancy (NV) center in fluorescent nanodiamonds (FNDs) has emerged as a good candidate to sense temperature, pH, and the concentration of paramagnetic species at the nanoscale, but comes with several hurdles such as particle-to-particle variation which render calibrated measurements difficult, and the challenge to tightly confine and precisely position sensors in aqueous environment. 
To address this, we demonstrate robust NV spin relaxometry within optically-trapped FNDs. 
In a proof of principle experiment, we show that optically-trapped FNDs enable highly reproducible nanomolar sensitivity to the paramagnetic ion, \(\mathrm{Gd}^{3+}\). 
We capture the three distinct phases of our experimental data by devising a model analogous to nanoscale Langmuir adsorption combined with spin coherence dynamics. Our work provides a basis for routes to sense free paramagnetic ions and molecules in biologically relevant conditions.

\end{abstract}

\date{\today}

\maketitle

\section{Introduction}
Investigating the electrical and magnetic properties of biologically relevant and aqueous solutions on the nanoscale has remained a persistent challenge. Progress here is expected to vastly improve our understanding of many complex biological processes such as electron transport in chemiosmosis, the generation of free radicals from redox reactions, and intracellular communication \cite{nie2021quantum, dai2023interface, barton2020nanoscale, rendler2017optical, fan2023diamond, barry2016optical, igarashi2020tracking, kuo2013fluorescent}.
Conventional techniques, such as fluorescent dyes, are limited by photobleaching and prone to artefactual changes to the signal.
The recent development of quantum sensors in fluorescent nanodiamonds (FNDs), namely the Nitrogen-Vacancy (NV) centers, offers a potential alternative. 
NV centers are atomic-scale defects whose spin levels are extremely sensitive to the local changes of temperature, pH, strain, and electromagnetic signals \cite{schrand2009nanodiamond, kucsko2013nanometre, fujiwara2020real, simpson2017non, petrini2022nanodiamond, hsieh2019imaging, Fujisaku2019, xu2024quantum,block2021optically, PeronaMartnez2020, bian2021nanoscale, aslam2023quantum, wu2022recent, rodgers2021materials, mochalin2020properties, shi2015single, choi2020probing, chen2022immunomagnetic, glenn2015single}. 
Moreover, the chemical inertness and high thermal conductivity make FNDs highly bio-compatible and highlight their ability to serve as nano-scale quantum sensors in biological conditions \cite{chipaux2018nanodiamonds,hsiao2016fluorescent,shenderova2015science, van2018nanodiamonds}. 

However, there are two major challenges that hinder the accurate use of FND-relaxometry in probing paramagnetic spins in an aqueous environment. 
The first challenge arises from the unaccounted charge state dynamics of NV centers, which make it difficult to quantitatively extract the spin relaxation timescale. 
Indeed, several prior studies have reported significant inconsistencies in the sensitivity of nanodiamonds to paramagnetic spin concentration, with values ranging from nanomolar to millimolar~\cite{PeronaMartnez2020,Vedelaar2023,Steinert2013,Tetienne2013}.
Such a wide range of reported sensitivity, in conjunction with the lack of a theoretical model to reconcile and connect the various results, lends to the difficulty in drawing quantitatively meaningful conclusions.
Second, due to random Brownian motion, positioning and detecting FND particles in solution remains a challenge.
Prior works have demonstrated the immobilization of FNDs to substrates through adhesion or functionalization~\cite{miller2020spin, jariwala2020surface,xie2022biocompatible}, as well as the use of an optical trap to spatially confine FNDs~\cite{Horowitz2012, hoang2016electron, Russell2018}.

In this work, we demonstrate robust spin-relaxometry with optically-trapped $70~$nm FNDs to sense free electronic spins in solution with nanomolar resolution.
Nanodiamonds form micrometer-sized aggregates due to optical trapping forces, facilitating the positioning and detection of FNDs in various solutions.
Using a robust differential measurement to eliminate the effects of charge dynamics, we reliably probe the spin relaxation time, $T_1$, of NV centers with a characteristic stretched exponential decay profile.
In a proof of principle experiment, we observe that the $T_1$ time of NV centers exhibits a novel triphasic response curve with the increasing concentrations of paramagnetic species, namely gadolinium 
\figone
We develop a comprehensive theoretical model to understand and quantitatively capture the three distinct regimes observed in experiment---surface dynamics, surface saturation, and solution dynamics.

\section{Results}

\emph{Characterization of Optically-Trapped Nanodiamond Sensors} ---
We use $70$~nm sized nanodiamond particles, each containing $\sim3~$part per million (ppm) NV centers (Fig.~\ref{fig:fig1}a).
Each NV center consists of a substitutional nitrogen impurity adjacent to a vacancy, replacing two intrinsic carbon atoms inside diamond.
The electronic ground state of NV centers exhibits a spin-1 degree of freedom.
In the absence of any external perturbations, $|m_s=\pm1\rangle$ spin sublevels are degenerate and separated from $|m_s=0\rangle$ by $2.87~$GHz.
These spin states can be optically initialized and read out, as well as coherently manipulated through microwave fields.

To realize 3-dimensional confinement and positioning of FNDs in an aqueous environment for sensing applications, we integrate a tightly focused, near-infrared laser trapping beam ($976$~nm) into our home-built confocal microscope \cite{supp}. We start with a sample chamber containing free FNDs suspended in deionized water (0.1 mg/mL).
When the trapping beam is off, the FNDs randomly diffuse in the solution, leading to a weak, near uniform fluorescence image as we scan the green excitation laser~(Fig.~\ref{fig:fig1}b).
In contrast, when the trapping beam is on, the dielectric nanodiamond particles experience $\sim 10~$pN of trapping force due to the scattering of photons and aggregate near the focus of the beam, resulting in a region with strong fluorescent signals.
Figure~\ref{fig:fig1}c shows an experiment where we monitor the fluorescence intensity at the center of the trapping beam while incrementally increasing power.
By mounting the sample chamber onto a piezo-electric stage, we further realize the 3D spatial control of a trapped FND aggregate in liquid solution.

The spin transition energies of NV centers can be probed using optically detected magnetic resonance (ODMR) spectroscopy: by sweeping the frequency of the applied microwave drive while monitoring the fluorescence signal, we expect a decrease in fluorescence when the microwave frequency is resonant with the electronic spin transition and drives the spin from $|m_s=0\rangle$ to the less bright $|m_s=\pm1\rangle$ sublevels.
Figure~\ref{fig:fig1}d displays the obtained ODMR spectrum from a trapped FND aggregate in water.
We observe the characteristic NV resonance at $2.87~$GHz with a small peak splitting, originating from the local strain and electric environment of the FNDs~\cite{mittiga2018imaging}.

Next, we use the NV spin's lifetime, \(T_1\), to sense the local magnetic fluctuation from paramagnetic spin species in a liquid environment.
To reliably probe the NV spin relaxation dynamics, we utilize a robust differential measurement scheme illustrated in Figure~\ref{fig:fig1}e. Specifically, after letting the NV centers reach charge state equilibrium for $100~\mu$s in the dark, we apply a $10~\mu$s green laser to initialize the spin state followed by free relaxation.
A second laser is applied at the end for fluorescence detection, with the photon counts designated as the bright signal, $S_\mathrm{B}(t)$.
By repeating the same sequence but with a final $\pi$-pulse before the readout to swap the spin populations between $|m_s=0\rangle$ and $|m_s=\pm1\rangle$, we measure the fluorescence of an orthogonal spin state to be the dark signal, $S_\mathrm{D}(t)$. The difference in fluorescence (contrast) between the two quantities can faithfully represent the measured spin relaxation dynamics of NV centers.
We remark that such a differential measurement scheme has been widely employed in studies of dense ensembles of solid-state spin defects to counter the photo-ionization process.
If one only accounts for the $S_\mathrm{B}(t)$ (a commonly used all-optical measurement scheme in prior nanodiamond relaxometry experiments), the measured dynamics can be highly dependent on the laser intensity and does not reflect the actual spin relaxation ($T_1$) process of the NV centers within FNDs \cite{mrozek2015longitudinal,choi2017depolarization,Gong2023}.

Interestingly, we find that the $T_1$ decay follows a characteristic stretched exponential profile, $\sim e^{-(t/T_1)^{0.5}}$~(Fig.~\ref{fig:fig1}e), rather than a conventional single exponential profile~\cite{choi2017depolarization, davis2023probing}.
This stretched exponential profile can be understood as the average effect from an ensemble of NV centers in FNDs, as each has a different decay timescale sensitive to the local environment (see the “Theoretical model" section). 
For optically-trapped $70~$nm FNDs in deionized water, we extract $T_1$ timescales ranging from $40-60~\mu$s.

\figtwo

\emph{Sensing Paramagnetic Species with Nanomolar Resolution} ---

Now with optically-trapped FNDs in hand, we then seek to probe paramagnetic species in aqueous environments using $T_1$ relaxometry.
We employ the use of GdCl\(_3\), which dissociates into Cl$^-$ and Gd$^{3+}$ ions when dissolved in water. Because of its seven unpaired electrons, the Gd$^{3+}$ species is highly paramagnetic and has been widely used in magnetic resonance  experiments~\cite{Tetienne2013,sushkov2014all,Radu2019,PeronaMartnez2020,Gao2023-hd}. 

To experimentally determine the effect of Gd$^{3+}$ on NV centers in optically-trapped FNDs, we collect measurements at a series of Gd$^{3+}$ concentrations spanning more than 9 orders of magnitude, from 1~nM to 5~M~(Figure~\ref{fig:fig2}a). 
FNDs are first suspended in deionized water and mixed with GdCl$_3$ solution; then this resulting solution is injected into a fluidics chamber.
We apply 93~mW of the IR trapping laser beam to confine and form a FND aggregate with micrometer-scale sizes,
and then perform ODMR and \(T_1\) measurements at the center of the aggregate.
We notice that the suspended FNDs tend to aggregate more at higher GdCl$_3$ density, consistent with prior studies finding that salts can diminish the repulsive forces between the negatively charged surfaces of FNDs ~\cite{hemelaar2017interaction}.
At each Gd\(^{3+}\) concentration, the $T_1$ measurement is repeated for seven different aggregates to obtain sufficient statistics to account for the particle-to-particle variation.
After acquiring a measurement, we flush the chamber with additional FND-Gd\(^{3+}\) solution to dissipate the trapped FND aggregate, allowing for a new aggregate to be formed in the trap.
We flush and clean the fluidics chamber with deionized water before moving on to a different Gd\(^{3+}\) concentration.
\figthree

Intuitively, paramagnetic spins in solution will generate magnetic fluctuations near the FNDs, leading to a reduction of the NV spin relaxation time.
This is indeed borne out of our data. 
As shown in Figure~\ref{fig:fig2}a, the measured NV $T_1$ drops from $\sim50~\mu$s in deionized water to $\sim15~\mu$s with $10~\mu$M Gd\(^{3+}\), and to $\sim7~\mu$s with $1~$M Gd\(^{3+}\).
Moreover, the reduction of $T_1$ at higher Gd\(^{3+}\) density is further corroborated with the decrease of ODMR contrast (Figure~\ref{fig:fig2}a Inset), as a shorter spin lifetime can result in worse optical initialization efficiency under green laser excitation.

Figure~\ref{fig:fig2}b summaries the dependence of NV $T_1$ on Gd\(^{3+}\) concentration.
Crucially, in contrast to a simple monotonic decay of $T_1$ with increasing [Gd\(^{3+}\)], we observe a clear triphasic response: the $T_1$ timescale first exhibits a sharp drop from 1~nM to 100~nM, and then plateau within a broad range of [Gd\(^{3+}\)], followed by another drop beyond $10~$mM concentration.
To confirm that the change in the $T_1$ indeed comes from paramagnetic species Gd\(^{3+}\), we perform another set of relaxometry experiments in solutions of NaCl, as neither Na$^+$ nor Cl$^-$ ions carry unpaired electrons.
In this situation, we observe that the $T_1$ is independent of NaCl concentration (Figure~\ref{fig:fig2}b Inset).

\emph{Theoretical model} ---
To capture the observed triphasic response of NV $T_1$, we develop a theoretical model accounting for both the freely moving paramagnetic spins in solution and the spins attracted towards the surface of the FNDs. In particular, the initial drop in $T_1$ at nanomolar concentrations can be understood through the attraction of Gd$^{3+}$ ions from the solution to the FND surface.
The FNDs used in this work have carboxyl surface (-COOH) groups which may confer a negative charge to the FND's surface \cite{Fujisaku2019}
(Fig.~\ref{fig:fig3}a inset). 
As positively charged Gd$^{3+}$ ions are introduced into the solution, some of them can be attracted to the FND surface via Coulomb interactions, leading to an effective shell of dense paramagnetic spins at the surface of the FNDs, even at nanomolar concentration.
These surface paramagnetic spins are responsible for the initial sharp drop of NV $T_1$ from 1~nM to 100~nM Gd$^{3+}$ concentration. 
As the concentration of Gd$^{3+}$ keeps increasing, the amount of surface spins saturates due to the limited availability of negatively charged bonds on the FND surface (Fig.~\ref{fig:fig3}b).
This results in the second phase where the $T_1$ response plateaus within a broad range from 100~nM to 10~mM.
When the concentration of Gd$^{3+}$ exceeds $10~$mM, the spins in the solution become the dominant source for magnetic noise, and the $T_1$ of NV centers continue to decrease in the third phase (Fig.~\ref{fig:fig3}c).
To achieve a quantitative agreement between our model and experimental data, we theoretically calculate the $T_1$ of NV centers in FNDs in the presence of Gd$^{3+}$ ions.
Within a single $70~$nm FND, there exists on average $N\approx 100$ NV centers, and the measured spin relaxation decay is the summation of $T_1$ times from all individual NVs.
Assuming the decay profile for each NV center, labeled by index $i$, follows a single exponential decay with timescale $T_{1,i}$, the measured $T_1$ takes the form, 
\begin{equation}
    C(t) = \frac{1}{N} \sum_{i=1}^{N} e^{-t/T_{1,i}}.
\end{equation}

For each NV center, $T_{1,i}$ is induced by the paramagnetic spins from the surrounding, whose value can be estimated following prior work~\cite{Tetienne2013},
\begin{equation}
    \Gamma_i = \frac{1}{T_{1,i}} = \sum_{j} 3 \gamma_e^2 B_{\perp,j}^2 \frac{\tau_{c,j}}{1+\omega^2 \tau_{c,j}^2}
\end{equation}
where $j$ refers to the paramagnetic spin in the environment, including both surface spins (characterized by surface density $\sigma$) and the spins in solution (characterized by volume density $\rho =$ [Gd$^{3+}$]), $B_{\perp,j}^2$ is the strength of the magnetic field at the site of the NV center (perpendicular to the NV axis), $\tau_{c,j}$ is the correlation time of the field, and $\omega=2.87~$GHz is energy difference between $|m_s=0\rangle$ and $|m_s=\pm1\rangle$.
By averaging across all possible random positions of NV centers within the FND, as well as the surrounding paramagnetic spins, we fit the theoretically calculated $T_1$ decay to the experimental data and extract values for the surface spin density, $\sigma$, as a function of volume spin density $\rho$. At a given $\rho$, the only fitting parameter in our model is $\sigma$, while all other terms can be independently estimated \cite{supp}.

Using our model, we successfully reproduce the characteristic stretched exponential $T_1$ decay profiles from experiment across all $[\mathrm{Gd}^{3+}]$ concentrations~(Fig.~\ref{fig:fig3}a). 
Such agreement further corroborates the feasibility of our theoretical model.
Figure~\ref{fig:fig3}b shows the extracted surface spin concentration value, $\sigma$, which saturates at around $8.4\times10^{17}~$m$^{-2}$, corresponding to an average spacing of $1.1~$nm between surface spins. We note that even in deionized water, there exists a finite density of surface spins from the surface dangling bonds.
This agrees with the previous studies finding that shallow NV centers typically exhibit a shorter $T_1$ timescale compared to NV centers in bulk diamond ~\cite{Tetienne2013, Jarmola2012,Dwyer2022}.
Combining both the contributions of spins from the surface and solution, our model proves to be in complete congruence with the experimentally measured $T_1$ timescale across all three stages (Fig.~\ref{fig:fig3}c).
The minor discrepancy between our model and the experimental data may arise from the variations in FND sizes, the surface ion saturation process, and the estimated correlation time of the Gd$^{3+}$ spins \cite{supp}.

To further validate our microscopic model of FND surface ion absorption, we perform additional spin-relaxometry experiment in solutions with a fixed $100$~mM NaCl~(Fig.~\ref{fig:fig3}c).
Here, the NaCl concentration is chosen to mimic the salt background observed in realistic subcellular environments. 
The Na$^{+}$ in solution can compete with Gd$^{3+}$ for occupancy of negatively charged groups on the FND surface, leading to a lower surface spin density $\sigma$ at a given volume density [Gd$^{3+}$].
This is indeed borne out by our data.
We observe that the NV $T_1$ starts to decrease at [Gd$^{3+}$]$\sim100~\mu$M rather than 1~nM.
As the [Gd$^{3+}$] continues to increase to around $10~$mM, the measured $T_1$ time approaches the original results in solution without NaCl.

\section{Outlook}

Our work opens the door to several intriguing future directions.
On the technological front, the nanomolar resolution of detection demonstrated here relies on the carboxylated surface of FND sensors to attract charged paramagnetic species from the solution.
One interesting direction to explore would be the potential to boost quantum sensing and fine-tune the sensitivity of FNDs using various methods of surface functionalization~\cite{xie2022biocompatible,janitz2022diamond,ackermann2019efficient, sangtawesin2019origins, kawai2019nitrogen, kayci2021multiplexed}.
On the scientific front, paramagnetic species regulate critical physiological processes, including metabolism and cell signaling~\cite{nie2021quantum, dai2023interface}. 
Achieving localized detection and quantification of these species will be essential to bolster the mechanistic understanding of these processes in living systems.
\vspace{2mm}

\vspace{1mm}

\emph{Acknowledgements}: We gratefully acknowledge assistance in the early stage of the experiment from R.~Gong, Z.~Liu, G.~He and X.~Du.
We thank A.~Jayich, Z.~Zhang and M.~Xie for helpful discussions.
This work is supported by the seed funding from the Center for Quantum Leaps at Washington University.
E.H. acknowledges support from the Gordon and Betty Moore Foundation, grant DOI 10.37807/gbmf11560.
S.M. acknowledges support from NIH grant R35GM142704.
C.Z. acknowledges support from NSF ExpandQISE 2328837.

\vspace{1mm}

\bibliographystyle{ieeetr}
\bibliography{ref}

\nocite{panich2011nuclear}
\nocite{simpson2017electron}
\nocite{bluvstein2019identifying}

\end{document}